\documentclass[amsmath,amssymb,prc,final,showpacs,twocolumn]{revtex4-1} 

\usepackage{bm,hyphenat,xspace} %,draftcopy}
\usepackage{graphicx,epsfig}

\newcommand{\A}[2]{{}^{#1}\mathrm{#2}}

\newcommand {\mbf}[1]{{\mathbf{#1}}}

\newcommand {\vecg}[1]{\mbox{\boldmath{$#1$}} }
\newcommand{\cm}{\mathrm{c\!\:\!.m\!\:\!.}}

\begin{document}

\title {Calculation of three-body nuclear reactions with angular-momentum
and parity-dependent optical potentials}

\author{A.~Deltuva} 
\email{arnoldas.deltuva@tfai.vu.lt}
\author{D. Jur\v{c}iukonis}
\affiliation{Institute of Theoretical Physics and Astronomy, 
Vilnius University, Saul\.etekio al. 3, LT-10222 Vilnius, Lithuania}
%Centro de F\'{\i}sica Nuclear da Universidade de Lisboa, 
%P-1649-003 Lisboa, Portugal }

%\received{October 11, 2015}


\begin{abstract}
Angular-momentum or parity-dependent nonlocal optical potentials
for nucleon-${}^{16}\mathrm{O}$ scattering able to fit 
differential cross section data over the whole angular regime are developed
and applied to the description of deuteron-${}^{16}\mathrm{O}$ scattering
in the framework of three-body Faddeev-type equations for transition operators.
Differential cross sections and deuteron analyzing powers for elastic scattering
and ${}^{16}\mathrm{O}(d,p){}^{17}\mathrm{O}$ transfer reactions are calculated
using a number of local and nonlocal optical potentials and compared with experimental data.
Angular-momentum or parity-dependence of the optical potential turns out to be quite irrelevant
in the considered three-body reactions while  nonlocality is essential
for a successful description of the differential cross section data,
especially in transfer reactions.
\end{abstract}

 \maketitle

%%%%%%%%%%%%%%%%%%%%%%%%%%%%%%%%%%%%%%%%%%%%%%%%%%%%%%%%%%%%%%%%%%%%%%%%%%%%%%%
\section{Introduction \label{sec:intro}}

Deuteron scattering from composite nuclei is, in general, a many-body
problem, but very often it is treated as three-body problem in a system
consisting of proton ($p$), neutron ($n$), and inert nuclear core $A$
\cite{johnson:70a,austern:87,deltuva:07d}.
The interactions between nucleons and core are modeled by
effective optical potentials (OPs); there is a large variety of
phenomenological parametrizations \cite{watson,menet,becchetti,ch89,koning}.
In some cases such models are successful in describing the
experimental data, and in some cases fail, calling for improvements
in the OP or the dynamical model, e.g., by including the excitation
of the core \cite{deltuva:13d}.
In particular cases the failures can be seen already in the nucleon-nucleus
two-body system. For example, standard optical potentials were
found to be unable to account for precise large-angle nucleon-nucleus 
elastic differential cross section data above 20 MeV lab energy for stable
closed-shell nuclei such as $\A{16}{O}$ or $\A{40}{Ca}$
\cite{kobos:79}.
This is not very surprising given significant momentum transfer in that
regime and possible excitations.
It was argued \cite{mackintosh:79,rawitscher:04} that these effects give rise to 
parity- ($\pi$) or angular-momentum ($L$) dependent components in the OP. 
Indeed, with such additional terms the description of large-angle differential 
cross section data was significantly improved
\cite{kobos:79,Cooper199787}.
Naturally one may raise the question what consequences the improvements
and new OP terms have on three-body reaction observables.
Unfortunately, explicitly $\pi$- or $L$-dependent OPs are not suitable for standard
practical calculations within distorted wave Born approximation (DWBA),
adiabatic wave approximation (ADWA) and continuum discretized coupled channels
 (CDCC) frameworks.
However, exact Faddeev-type theory, implemented in individual 
partial-wave representations for all three involved pairs,
is capable of using such $\pi$- and $L$-dependent potentials.
Thus, the aim of the present work is to study three-body nuclear reactions
with $\pi$- and $L$-dependent OPs. We choose $d+\A{16}{O}$ elastic and
transfer reactions as a working example. However, all existing
$\pi$- and $L$-dependent OP parametrizations for nucleon-$\A{16}{O}$ 
are local \cite{kobos:79,Cooper199787}, 
while in Ref.~\cite{deltuva:09b} it was found that the Perey-Buck-type exchange 
nonlocality of the OP is important in three-body reactions, especially for transfer.
For this reason we create several nonlocal parametrizations of
$\pi$- and $L$-dependent OPs, fitted to the same experimental data \cite{p16o:sexp,p16o:aexp}
as the local ones \cite{kobos:79,Cooper199787}.

In Sec.~\ref{sec:2} we describe the employed two-body
nucleon-$\A{16}{O}$ potentials, and in Sec.~\ref{sec:3}
the three-body scattering equations.
Results are presented in  Sec.~\ref{sec:4},
while summary is given in  Sec.~\ref{sec:5}.

\section{Nucleon-nucleus potentials \label{sec:2}}
\subsection{Angular-momentum-dependent optical potential}

\begin{figure}[!]
\begin{center}
\includegraphics[scale=0.59]{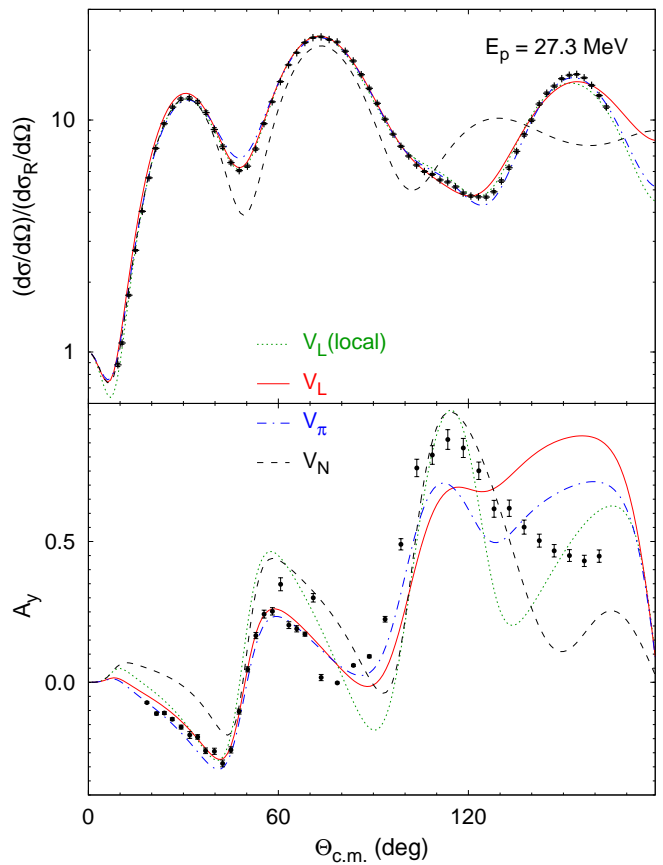}
\end{center}
\caption{\label{fig:po}  (Color online)
Differential cross section divided by Rutherford cross section
and proton analyzing power for  $p+\A{16}{O}$ elastic scattering at $E_p = 27.3$ MeV. 
Parameters of local $L$-dependent OP from  Ref.~\cite{kobos:79} (dotted curves),
nonlocal  $L$-dependent OP (solid curves) and nonlocal $\pi$-dependent OP (dashed-dotted curves)
are fitted to experimental  $E_p = 27.3$ MeV data. Results for  nonlocal $L$-independent OP
(dashed curves) without proper fit are shown as well.
The  data are from Refs.~\cite{p16o:sexp,p16o:aexp}.}
\end{figure}

\begin{figure}[!]
\begin{center}
\includegraphics[scale=0.5]{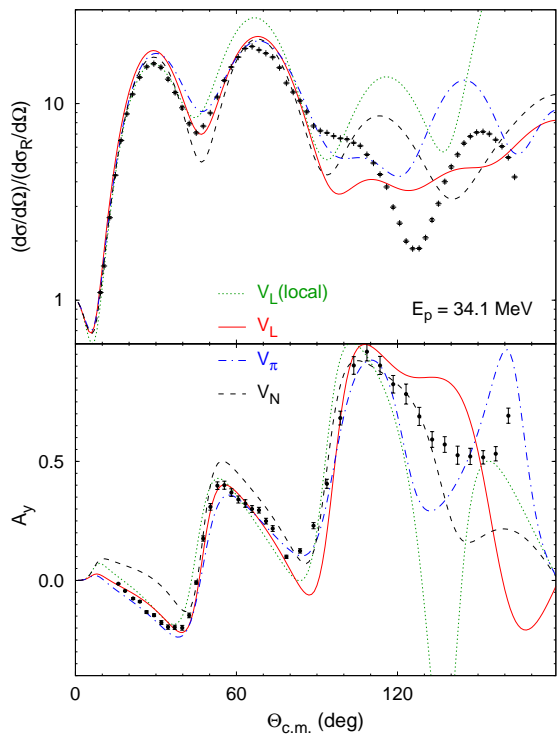}
\end{center}
\caption{\label{fig:po34}  (Color online)
Differential cross section divided by Rutherford cross section
and proton analyzing power for  $p+\A{16}{O}$ elastic scattering at $E_p = 34.1$ MeV. 
Curves are as in  Fig.~\ref{fig:po} with OP parameters fitted to experimental 
$E_p = 27.3$ MeV data. The results are not constrained in any way by the
shown $E_p = 34.1$ MeV data from Refs.~\cite{p16o:sexp,p16o:aexp}.}
\end{figure}

We start with a nonlocal OP of the form proposed by Giannini and Ricco \cite{giannini}
and augment it with an  $L$-dependent part, resulting 
\begin{gather}  \label{eq:VL}
\begin{split} \
V_{L}(\mbf{r}',\mbf{r}) =  {}& - H_c(x)[V_V \,f_V(y) + iW_V \,f_W(y) + i W_S\,g_S(y)]  \\ {}&
- H_s(x) V_s \frac{2}{y} \frac{df_s(y)}{dy} \, \vecg{\sigma}\cdot \mbf{L} \\ {}&
- H_c(x)[\tilde{V}\,g_{\tilde{V}}(y) + i \tilde{W}\,g_{\tilde{W}}(y)] \, f_L(L^2).
\end{split}
\end{gather}
Here $V_i$ and $W_i$ are potential strengths for various real and imaginary terms, 
 while for each term the shape is given by
\begin{subequations}  \label{eq:VLdef}   
\begin{align} 
x = {}& |\mbf{r}'-\mbf{r}|, \\
y = {}& |\mbf{r}'+\mbf{r}|/2, \\ 
H_i(x) = {}& (\pi \beta_i^2)^{-3/2} e^{-x^2/\beta_i^2}, \\
f_i(y) = {}& [1+ e^{(y-R_i)/a_i}]^{-1},  \label{eq:fws} \\
g_i(y) = {}& -4a_i df_i(y)/dy = 4f_i(y)[1-f_i(y)]
\end{align}
\end{subequations}
with radius $R_i$, diffuseness $a_i$, and nonlocality parameter $\beta_i$.
The dependence on the orbital angular momentum $L$ in the form of 
Woods-Saxon function  $f_L(L^2)$ given in Eq.~(\ref{eq:fws}) is taken over
from Ref.~\cite{kobos:79} where local $L$-dependent potential was constructed.
We determine potential  strength parameters by fitting theoretical $p+\A{16}{O}$ predictions
to experimental data for the differential cross section $d\sigma/d\Omega$, 
total inelastic cross section, and proton
analyzing power $A_y$ from Refs.~\cite{p16o:sexp,p16o:aexp}, i.e., same data that constrain 
the local OP of Ref.~\cite{kobos:79}.
In addition, also geometric parameters $R_i$, $a_i$, and  $\beta_i$ have been varied, 
typically within 10\% of original values  \cite{giannini}, to improve the fit quality.
The parameters defining the $L$-dependence turn out to be comparable to the 
local case \cite{kobos:79}, i.e., typically $3 < \sqrt{R_L} < 4$ and $0.5 < \sqrt{a_L} < 0.7$.
Note that some local OP parameters in Ref.~\cite{kobos:79} depend strongly on the collision
energy and that dependence is not smooth, reflecting the fact that
backward-angle experimental data exhibit non-monotonic energy dependence,
probably due to the presence of resonant proton-nucleus states. 
We emphasize that in three-body reactions the energy of each interacting two-body 
subsystem formally runs from the available three-body energy to $-\infty$, 
but, in order to have a single three-body Hamiltonian
and thereby preserve a Hamiltonian theory \cite{deltuva:09a},
  it is preferable to use two-body potentials with fixed sets of parameters; 
the results in Sec.~\ref{sec:3} are obtained following this strategy.
The ability of the OP  to account for the two-body reaction data over a 
broader energy regime may be important for its success in three-body
reactions and deserves investigation.
The nonlocal OP, at least to some extent, is able to absorb smooth energy dependence of data into nonlocality,
but far less a non-smooth behavior. Thus, we have not achieved a single parameter set
describing the data over the full angular regime at all energies. In fact,
the analyzing power data are only accounted for  center-of-mass (c.m.)
scattering angles $\Theta_\cm$ up to about $100^\circ$.
Nevertheless, when we fit the experimental data at a given energy, the resulting OP
reproduces well the data for other energies at  $\Theta_\cm \le 90^\circ$
and only fails at backward angles where the differential cross section is very small.
This is an important improvement as compared to the local OP of Ref.~\cite{kobos:79}
that yields considerably worse description for the data not included in the fit.
An example is shown in Figs.~\ref{fig:po} and \ref{fig:po34} where the predictions 
using local and nonlocal $L$-dependent OPs, 
determined solely by the data at proton lab energy $E_p= 27.3$ MeV, 
are compared with data at  $E_p= 27.3$ and 34.1 MeV. Thus, Fig.~\ref{fig:po} reflects
the quality of the fit at a single given energy, while Fig.~\ref{fig:po34} 
reflects the predictive power of fixed-energy  OPs for 
energies not included in the fit. In this latter case  nonlocal OPs are more successful,
indicating weaker energy dependence of their parameters as compared to the local OP.
We present also predictions of parity-dependent OP from the next subsection, $V_\pi$, and
of nonlocal $L$-independent OP with parameters from
Ref.~\cite{deltuva:09b}, labeled $V_N$. The latter was not properly fitted to the present data
failing at backward angles and, for $A_y$, also at forward angles  $\Theta_\cm \le 40^\circ$,
but otherwise provides a rough description of the experimental data.
To confirm the conclusion on the superiority of nonlocal $L$-dependent OP in the
two-body system, we created a number of parametrizations \cite{kiti}.
%%%%%
%%%%%

\begin{figure}[!]
\begin{center}
\includegraphics[scale=0.7]{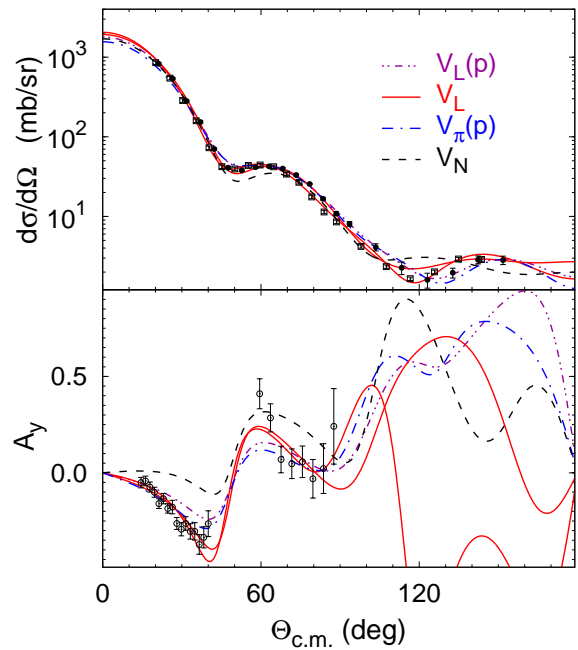}
\end{center}
\caption{\label{fig:no}  (Color online)
Differential cross section and neutron analyzing power for  $n+\A{16}{O}$ elastic 
scattering at $E_n = 24$ MeV. Results obtained with various nonlocal potentials 
are compared with the experimental data from Refs.~\cite{Grabmayr1980167} ($\bullet$),
\cite{delaroche:86} ($\Box$), and \cite{lam:85} ($\circ$).
The predictions with $p+\A{16}{O}$ potential parameters from Fig.~\ref{fig:po} are
denoted by (p). }
\end{figure}

As for  $n+\A{16}{O}$ scattering, the available experimental data 
\cite{Grabmayr1980167,lam:85,delaroche:86}
are scarcer and
less precise. We tried two options for  $L$-dependent $n+\A{16}{O}$ potential:
i) taking over the parameters of the $p+\A{16}{O}$ potential,
ii) explicitly fitting to $n+\A{16}{O}$ experimental data. 
An example for $n+\A{16}{O}$ scattering at $E_n=24$ MeV neutron energy is presented in Fig.~\ref{fig:no}.
While for the differential cross section the quality is nearly the same in both cases,
explicit fitting leads to a better description of the neutron analyzing power. 
We show results for two sets of parameters to demonstrate large model dependence
for backward angle $A_y$.

%%%%%

\subsection{Parity-dependent optical potential}

The mechanism giving rise to  parity-dependent terms in the OP 
\cite{rawitscher:04,Cooper199787,Cooper1995338}.
is different from that of $L$-dependent terms, but nevertheless has led to a 
comparable quality when fitting the data. To get the nonlocal version we
replace the $f_L(L^2)$ term in Eq.~(\ref{eq:VL}) by a parity-dependent term,
 resulting in a nonlocal $\pi$-dependent OP
\begin{gather}  \label{eq:VLpi}
\begin{split} 
V_{\pi}(\mbf{r}',\mbf{r}) =  {}& - H_c(x)[V_V \,f_V(y) + iW_V \,f_W(y) + i W_S\,g_S(y)]  \\ {}&
- H_s(x) \frac{2}{y} \left[V_s  \frac{df_s(y)}{dy} + iW_s\frac{df_{s_i}(y)}{dy} \right] 
\, \vecg{\sigma}\cdot \mbf{L} \\ {}&
- (-1)^L H_c(x) [\tilde{V}_V \,f_{\tilde{V}}(y) + i \tilde{W}_V \,f_{\tilde{W}}(y) \\ {}& +
\tilde{V}_S\,g_{\tilde{S}_r}(y) + i \tilde{W}_S\,g_{\tilde{S}_i}(y) ].
\end{split}
\end{gather}
In addition, we allowed for an imaginary spin-orbit term  with strength $W_s$, following
Ref.~\cite{Cooper199787} where  $\pi$-dependent local OP was developed, but fixed
nonlocality parameters $\beta_i$ to original values from Ref.~\cite{giannini}.
Otherwise the fitting procedure is the same as in previous subsection, and the achieved quality
in describing experimental data is comparable to that of $L$-dependent OP as can be seen in 
 Fig.~\ref{fig:po}.
Worth noting are different radial shapes for  $\pi$-dependent and independent terms,
in particular significantly smaller diffuseness $a_{\tilde{V}}$ as compared
to $a_V$, consistently
with findings of Ref.~\cite{rawitscher:04}.

\subsection{Discussion}
$L$- and $\pi$-dependence of the OP is not really surprizing, since OP describes not a
fundamental interaction but an effective one between the nucleon and composite nucleus.
Internal degrees of freedom of the nucleus $A$ , if taken into account explicitly, would
lead to a highly complicated effective two-body nucleon-nucleus interaction.
Solving the $(A+1)$-nucleon scattering problem with sufficient accuracy is beyond the
present capabilities, but attempts have been made to justify the nonstandard OP terms by
the effect of simplest internal degrees of freedom of the nuclear core.
E.g., Ref.~\cite{rawitscher:04} argues that contributions of the core excitation can be
approximated by  $\pi$-dependent terms, while  Refs.~\cite{mackintosh:79,kobos:79} relate
the $L$-dependence of the OP to the deuteron channel coupling. For curiosity we verified
this concept in a toy model using theoretical results from Ref.~\cite{deltuva:09b}  
for proton elastic scattering on $\A{16}{O}$ and $\A{17}{O}$, but we expect it to be
qualitatively valid for any nuclei $(A-1)\equiv B$ and $A$.
Starting with an  $L$-independent OP for  $p+B$, a real binding potential for  $n+B$, 
and a realistic $n+p$ potential having central, spin-spin, spin-orbit, and tensor terms,
results for  $p+A$ were obtained by solving exact
three-body equations, thereby including $p+n+B$ breakup and $d+B$
transfer channels to all orders. The resulting three-body $p+A$ elastic cross section 
could not be fitted well with the two-body standard OP for $p+A$,
but the inclusion of $L$-dependent terms in the  OP for $p+A$, i.e.,
$p+\A{17}{O}$ in case of Ref.~\cite{deltuva:09b},  significantly improved the fit.
Obviously, such an approach is not reliable for a quantitative determination of the OP
as it takes into account only one-neutron internal degrees of freedom in the core 
and relies on the potentials for the $p+B$ and $n+B$ subsystems
(that may be $L$-dependent themselves), but it demonstrates that the 
$L$-dependence of the OP appears. Of course, $L$-independent phenomenological OPs
do not exclude the coupling to the deuteron channel, but include it implicitly in an
$L$-averaged way by fiting the data.
%, althouhg not  the one of Ref.~\cite{kobos:79}.
Applying an $L$-dependent potential in a three-body $p+n+A$ system with present deuteron channel
is justified in exact calculations dealing with three pairwise 
$p+A$, $n+A$, and $n+p$ interactions,
but may lead to double counting in DWBA-type approaches that generate the
 $p+(A+n)$ wave not through a rigorous solution of the hree-body problem but
from a  $p+(A+n)$ two-body OP.
We also expect that if one would attempt to calculate
deuteron-nucleus two-body OP starting from a three-body problem with
$L$-independent nucleon-nucleus OPs, the resulting OP in a similar way should acquire 
$L$-dependence.

\section{Three-body scattering equations  \label{sec:3}}

We describe deuteron-nucleus reactions in the framework of exact Faddeev-type three-body
scattering equations. We use Alt-Grassberger-Sandhas (AGS) integral equations \cite{alt:67a}
for  transition operators 
\begin{equation}  \label{eq:Uba}
U_{\beta \alpha}  = \bar{\delta}_{\beta\alpha} \, G^{-1}_{0}  +
\sum_{\gamma}   \bar{\delta}_{\beta \gamma} \, T_{\gamma} 
\, G_{0} U_{\gamma \alpha},
\end{equation}
with $ \bar{\delta}_{\beta\alpha} = 1 - \delta_{\beta\alpha}$,  free resolvent
$G_0(Z) = (E+i0-H_0)^{-1}$, three-particle relative motion kinetic energy operator $H_0$
and available energy $E$,
and two-body transition matrix
\begin{equation}  \label{eq:T}
T_{\gamma} = v_{\gamma} + v_{\gamma} G_0 T_{\gamma}.
\end{equation}
The latter is calculated for each pair  $\gamma$ with
the corresponding  two-body potential $v_{\gamma}$, where in the 
 odd-man-out notation $v_{1}$ denotes the interaction within the pair (23) and so on.
On-shell matrix elements of $U_{\beta \alpha}$ taken between the corresponding channel states
are reaction amplitudes needed for the calculation of scattering observables.

We solve AGS equations in the momentum-space partial-wave representation.
We employ three complete sets of base functions
$|p_\alpha q_\alpha 
(l_\alpha \{ [ L_\alpha (s_\beta s_\gamma)S_\alpha] j_\alpha s_\alpha\} \mathcal{S}_\alpha)
J M \rangle $.
Here $(\alpha\beta\gamma) = (123)$, $(231)$, or $(312)$,
$p_\alpha$ is magnitude of relative momentum within pair  $(\beta\gamma)$,  
  $q_\alpha$ is magnitude of relative momentum between spectator $\alpha$ and 
c.m. of pair  $(\beta\gamma)$, $L_\alpha$ and  $l_\alpha$ are orbital
angular momenta associated with  $|p_\alpha$ and $ q_\alpha$, respectively,
and $s_\alpha,s_\beta, s_\gamma$ are spins of the corresponding particles.
All discrete angular momentum quantum numbers are coupled to total
angular momentum $J$ with the projection $M$,
while $S_\alpha$, $j_\alpha$ and $ \mathcal{S}_\alpha$ are angular momenta
 of intermediate subsystems.
Using all three sets $\alpha=1$, 2, and 3 of these basis states allows
the calculation of each potential $v_\alpha$ and transition matrix 
$T_\alpha$ in its proper basis. Obviously, this enables easy inclusion 
of $L$- and $\pi$-dependent potentials, in contrast to CDCC and other approximations,
where only one set of base functions is being used.

The proton-nucleus Coulomb force is included via the screening and renormalization
method \cite{taylor:74a,alt:80a,deltuva:05a}. For $d+\A{16}{O}$ elastic  and
transfer reactions we obtain well-converged results with Coulomb
screening radius around 10 or 12 fm, and including $J\le 30$ states
with $L_\alpha$ up to 3, 8, and 14 for $n+p$, $n+\A{16}{O}$, and $p+\A{16}{O}$
pairs, respectively. For the $n+p$ interaction we take the realistic
CD Bonn potential \cite{machleidt:01a} and use  potentials from previous section
for nucleon-nucleus pairs.

\section{Results \label{sec:4}}

\begin{figure}[!]
\begin{center}
\includegraphics[scale=0.71]{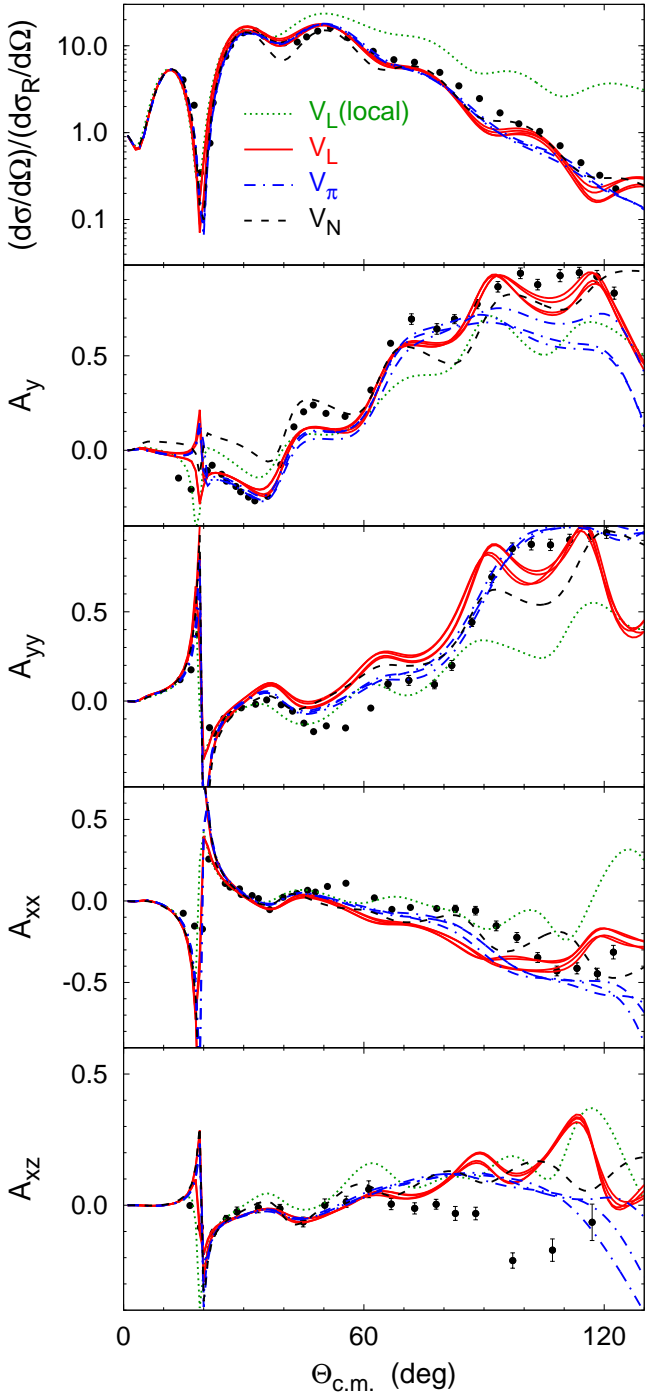}
\end{center}
\caption{\label{fig:do}  (Color online)
Differential cross section divided by Rutherford cross section
and deuteron analyzing powers for  $d+\A{16}{O}$ elastic scattering at $E_d = 56$ MeV.
 Predictions obtained using local $L$-dependent OP from  Ref.~\cite{kobos:79} (dotted curves),
four parametrizations of nonlocal  $L$-dependent OP (four solid curves,) 
three parametrizations of nonlocal $\pi$-dependent OP (three dashed-dotted curves),
and  nonlocal $L$-independent OP (dashed curves)
are compared with the experimental data from Ref.~\cite{matsuoka:86}.}
\end{figure}

Using  nucleon-nucleus OPs from Sec.~\ref{sec:2} and Refs.~\cite{kobos:79,giannini} and 
the realistic neutron-proton
CD Bonn potential \cite{machleidt:01a} we study 
$d+\A{16}{O}$ elastic scattering and transfer reactions 
$\A{16}{O}(d,p)\A{17}{O}$.
In the former case there exist differential cross section 
and deuteron analyzing power data at $E_d = 56$ MeV deuteron lab energy
\cite{matsuoka:86}.
Comparison of those experimental data and our predictions, including four 
 $L$-dependent and three $\pi$-dependent nonlocal models,
is presented in Fig.~\ref{fig:do}. One can notice immediately
that the local $L$-dependent OP from Ref.~\cite{kobos:79},
although being successful in $p+\A{16}{O}$ scattering,
fails heavily at large angles in $d+\A{16}{O}$ scattering,
strongly overpredicting the differential cross section.
In contrast, nonlocal models, both  $L$- or $\pi$-dependent, slightly
underpredict the differential cross section at large angles, but quite
reasonably follow its shape.
Properly fitted  $L$- and $\pi$-dependent models, both local and nonlocal, provide
a reasonable description of deuteron analyzing powers up to $\Theta_\cm = 40^\circ$
or  $60^\circ$ (in some cases, with exception of  $\Theta_\cm = 20^\circ$  
where  $d\sigma/d\Omega$ has a deep minimum), but deviate from data and from each other
at larger scattering angles. 
The $L$-independent nonlocal OP \cite{giannini} accounts for cross section data 
with a quality comparable to nonlocal $L$- and $\pi$-dependent OPs,
 but fails for the deuteron vector analyzing power
$A_y$ at  $20^\circ \le \Theta_\cm \le 40^\circ$. This is expected given that is
was not fitted  to $p+\A{16}{O}$  $A_y$ data and shows there a similar discrepancy.
Quite surprisingly, the description of all measured deuteron tensor analyzing powers
$A_{yy}$, $A_{xx}$, and $A_{xz}$ using this model \cite{giannini}
turns out to be quite similar to  $L$- and $\pi$-dependent models of Sec.~\ref{sec:2}
 and Ref.~\cite{kobos:79}. This may indicate that deuteron tensor analyzing powers
are not well constrained by nucleon-nucleus $A_y$ data.

\begin{figure}[!]
\begin{center}
\includegraphics[scale=0.69]{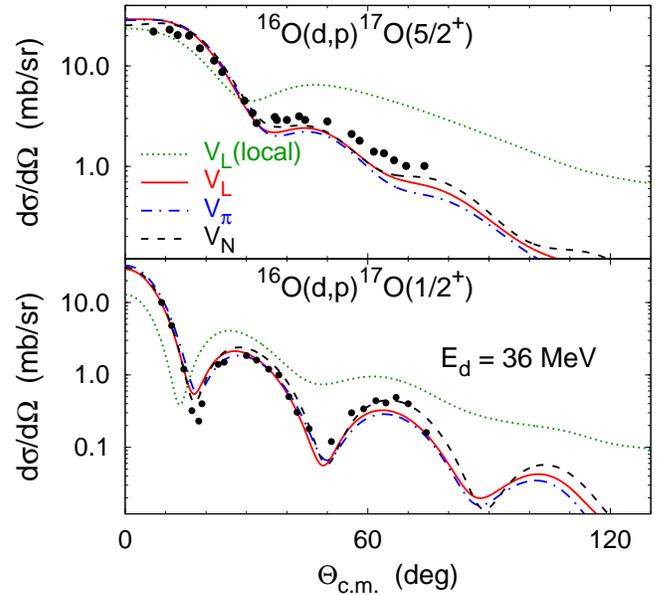}
\end{center}
\caption{\label{fig:dop}  (Color online)
Differential cross section  for  $\A{16}{O}(d,p)\A{17}{O}$ transfer reactions
at  $E_d = 36$ MeV leading to $\A{17}{O}$ ground $\frac52^+$ (top) and
excited $\frac12^+$ (bottom) states.
 Predictions obtained using local $L$-dependent OP from  Ref.~\cite{kobos:79} (dotted curves),
nonlocal  $L$-dependent OP (solid curves,) nonlocal $\pi$-dependent OP (dashed-dotted curves),
and  nonlocal $L$-independent OP (dashed curves)
are compared with experimental data from Ref.~\cite{dO25-63}.}
\end{figure}

\begin{figure}[!]
\begin{center}
\includegraphics[scale=0.69]{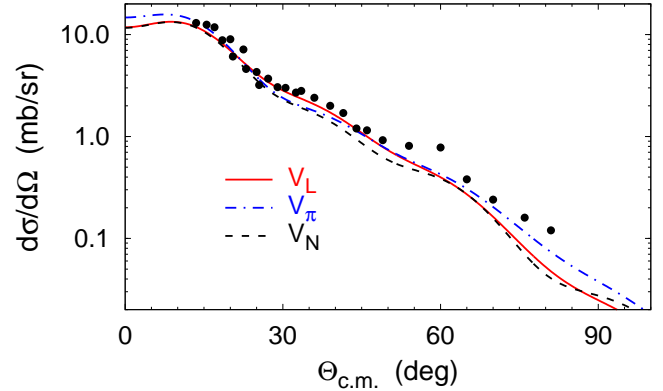}
\end{center}
\caption{\label{fig:dop63}  (Color online)
Differential cross section  for  $\A{16}{O}(d,p)\A{17}{O}$ transfer to
$\A{17}{O}$ ground state $\frac52^+$ at  $E_d = 63.2$ MeV.
 Predictions obtained using
nonlocal  $L$-dependent OP (solid curves,) nonlocal $\pi$-dependent OP (dashed-dotted curves),
and  nonlocal $L$-independent OP (dashed curves)
are compared with experimental data from Ref.~\cite{dO25-63}.}
\end{figure}

Next we study $\A{16}{O}(d,p)\A{17}{O}$ transfer reactions. In this case the potential
must support bound state for the final nucleus $ \A{17}{O}$. We therefore take real 
 binding potentials from Ref.~\cite{deltuva:15f} in  $n+\A{16}{O}$ partial wave
$\frac52^+$ ($\frac12^+$) when calculating transfer to  $ \A{17}{O}$ ground state
(excited state). Differential cross section results for both reactions at $E_d = 36$ 
MeV are shown in Fig.~\ref{fig:dop}.
By comparing with the experimental data \cite{dO25-63} one notices 
the failure of the local $L$-dependent OP from Ref.~\cite{kobos:79}
over the whole angular regime. In contrast, all nonlocal OPs,
$L$- or $\pi$-dependent or not, provide quite good description of the 
experimental data. Thus, $L$- and $\pi$-dependent terms in the OP appear to be
quite irrelevant while the nonlocality of the OP turns out to be essential.
Similar findings regarding the OP nonlocality in transfer reactions 
emerged in Refs.~\cite{deltuva:09b,deltuva:15f} where a broader range of reactions
was investigated. Note that there is a difference between present calculations
and those of in Ref.~\cite{deltuva:09b} in the choice of $n+\A{16}{O}$ potential:
it was real in Ref.~\cite{deltuva:09b} but complex here (except for $ \A{17}{O}$
bound state partial wave).
In  Fig.~\ref{fig:dop63} we present one more example, i.e., 
$\A{16}{O}(d,p)\A{17}{O}$ transfer to $ \A{17}{O}$ ground state $\frac52^+$ 
at  $E_d = 63.2$ MeV, not considered in Refs.~\cite{deltuva:09b,deltuva:15f}.
Again, the account of the experimental data \cite{dO25-63} by all employed
nonlocal potentials is quite good, while the calculations of
 Ref.~\cite{deltuva:09a} with local but explicitly energy-dependent potentials
heavily failed in reproducing this observable.

\section{Summary and conclusions \label{sec:5}}

We developed a number of angular-momentum or parity-dependent optical potentials
for nucleon-$\A{16}{O}$ system. Those nonstandard additional terms enabled
to fit elastic nucleon-nucleus scattering data at large angles.
However, the parameters turn out to be energy-dependent. Optical potential
with energy-independent parameters is able to fit two-body data very well around the chosen 
energy in the whole angular regime for the differential cross section and up to about
 $\Theta_\cm = 100^\circ$ for the analyzing power. At more distant energies the description remains
good at not too large scattering angles $\Theta_\cm \le 100^\circ$.
The local  $L$-dependent OP  from Ref.~\cite{kobos:79} turns out to be much stronger
energy-dependent, with a fixed parameter set able to account for the data in narrow energy
region only. In this respect the potentials of the present work represent a significant improvement.

The explicit angular-momentum or parity  dependence of the OP can be handled in the 
Faddeev/AGS three-body scattering equations solved in the momentum-space partial-wave
representation where each two-body potential and the corresponding transition matrix
is calculated in its proper basis.
In an energy-independent form the nonlocal  $L$- or $\pi$-dependent potentials 
were used to calculate differential cross section and  deuteron analyzing powers
for $d+\A{16}{O}$ elastic scattering
and $\A{16}{O}(d,p)\A{17}{O}$ transfer reactions.
To isolate the effect of  $L$- or $\pi$-dependence and nonlocality, same observables
were calculated using local  $L$-dependent \cite{kobos:79} and nonlocal 
$L$-independent \cite{giannini} potentials.
In all considered reactions nonlocal OPs provide quite similar and reasonable
description of differential cross section data. In contrast, the predictions using
the local  $L$-dependent OP \cite{kobos:79} strongly deviate from the data and
all nonlocal OPs for  $\Theta_\cm \ge 50^\circ$ in $d+\A{16}{O}$ elastic scattering
and in the whole angular regime for  $\A{16}{O}(d,p)\A{17}{O}$ transfer reactions.
Based on this fact we conclude that $L$- and $\pi$-dependent terms in the OP 
may be quite irrelevant for three-body scattering but the nonlocality plays a major role,
especially in transfer reactions;
the latter finding is in accordance with 
Refs.~\cite{deltuva:09b,deltuva:15f,timofeyuk:13b,titus:16a}.
The comparison of predictions and data for deuteron  analyzing powers
in $d+\A{16}{O}$ elastic scattering is less conclusive. The agreement is reasonable
for all properly fitted models at not too large  scattering angles  $\Theta_\cm \le 60^\circ$,
but beyond the predictions may deviate from data and from each other.
Furthermore, a proper fit to two-body  analyzing power
data appears to be relevant for deuteron vector analyzing power $A_y$, but not 
for  tensor analyzing powers $A_{yy}$, $A_{xx}$, and $A_{xz}$.

%\vspace{2mm}

This work was supported by Lietuvos Mokslo Taryba
(Research Council of Lithuania) under 
Contract No.~MIP-094/2015.

\vspace{1mm}
%\begin{acknowledgments}
%\end{acknowledgments}

%\clearpage

%%%%%%%%%%%%%%%%%%%%%%%%%%%%%%%%%%%%%%%%%%%%%%%%%%%%%%%%%%%%%%%%%%%%%%%%%%%%%
%\bibliographystyle{prsty}
%\bibliography{abbrev,pre80,80-89,90-99,200x,clmb,ad,nreact}

\begin{thebibliography}{10}
\bibitem{kiti}
Parameter tables and computer codes can be obtained from authors upon request.
\end{thebibliography}

\end{document}